\documentclass[runningheads]{svmult}

\usepackage{makeidx}   
\usepackage{graphicx}  
\usepackage{subeqnar}  
\usepackage{multicol}  
\usepackage{physprbb}  
\makeindex             

\begin{document}
\title*{Anti-Hierarchical Growth of Supermassive Black Holes and
  QSO lifetimes}
\toctitle{Anti-Hierarchical Growth of Supermassive Black Holes and QSO
  lifetimes}
\titlerunning{Anti-Hierarchical Growth of Supermassive Black Holes...}
\author{Andrea Merloni}
\authorrunning{Andrea Merloni}
\institute{Max-Planck-Institut f\"ur Astrophysik,
Karl-Schwarzschild-Strasse 1, D-85741, Garching, Germany}

\maketitle 

\begin{abstract}
 I present a new method to unveil the history of cosmic accretion and
 the build-up of Supermassive Black Holes (SMBH) 
in the nuclei of galaxies, based on observations 
of the evolving radio and (hard) X-ray luminosity functions of AGN. 
The fundamental plane of black hole activity discovered by 
Merloni, Heinz \& Di Matteo (2003) is used as a mass and accretion
 rate estimator. I adopt the local BH mass function as a boundary
 condition to integrate backwards in time the continuity equation for
 the SMBH evolution, neglecting the role of mergers. Under the 
most general assumption that accretion proceeds in a radiatively 
efficient way above a certain rate, and in a radiatively inefficient 
way below, the redshift evolution of the mass and accretion rate 
functions are calculated self-consistently. 
The only tunable parameters are the accretion efficiency and the 
critical ratio of the X-ray to Eddington luminosity at which the 
transition between accretion modes takes place. 
The evolution of the BH mass function between $z=0$ and $z \sim 3$ 
shows clear signs of an anti-hierarchical behaviour: while the 
majority of the most massive objects ($M > 10^9$) were already 
in place at $z \sim 3$, lower mass ones mainly grew at progressively 
lower redshift. As an example, I will discuss the consequences of
 these results for the lifetimes of accreting black holes.
\end{abstract}

\section{Introduction}
It has been known for the last ten years that the cosmological
evolution of massive galaxies shows signs of `down-sizing'
\cite{cow96}, i.e. of a progressive decrease of the typical 
mass of actively star forming galaxies. Many pieces of evidence,
brought forward also during this meeting (see e.g. the contributions
of Bender, Kauffmann and Danese to these Proceedings), 
suggest that the baryonic physics of star
formation processes counterbalance the hierarchical (bottom-up) 
growth of dark
matter structures, favouring the early formation and growth of the
most massive (and dense) stellar aggregates. 

The ubiquity of SMBH in galactic nuclei,
and their tight relation with their hosts' bulge properties \cite{mag}
seem to indicate that the formation and growth of galaxies and of
their resident black holes have proceeded in parallel, and probably
influenced each other in the early stages of galaxy formation. 
As a matter of fact, the number of theoretical  studies dedicated to 
AGN feedback in galaxy formation 
has flourished in the last five years (see e.g. Loeb's
contribution in this proceedings, and references therein).
Furthermore, a recent comprehensive
study of 23 000 local AGN carried out by the Sloan Digital Sky Survey
(SDSS, \cite{hec04}) have demonstrated, in a direct quantitative way, 
that accretion onto SMBH and formation of stars are tightly coupled
even in the local universe. 

Is it possible to learn more about the formation and growth of
structures by just looking at the evolution of the AGN population?
The aim of the work presented here is to show to what extent this is
indeed possible, and to describe a robust, self-consistent
way to unveil the history of cosmic accretion and
 the build-up of SMBH in the nuclei of galaxies, in the form of their
 evolving mass function. The methodology and the main results will be
 discussed in the next section, while in section~\ref{sec:times} I
 will  expand on the consequences of these results for the issue of
 QSO lifetimes. Section 4 will summarize my conclusions.

\section{The Evolution of the SMBH Mass Function}
\label{sec:res}

Under the standard assumption that black holes grow mainly by
 accretion \cite{yt02}, the cosmic evolution of the SMBH accretion rate and its
 associated mass density can be calculated from the luminosity
 function of AGN: $\phi(L_{\rm bol},z)=dN/dL_{\rm bol}$, where $L_{\rm
 bol}=\epsilon_{\rm rad} \dot M c^2$ is the bolometric luminosity produced by a
 SMBH accreting at a rate of $\dot M$ with a radiative efficiency
 $\epsilon_{\rm rad}$. In practice, the accreting black hole population is
 always selected through observations in specific wavebands.  Crucial
 is therefore the knowledge of two factors: the completeness of any
 specific AGN survey, and the bolometric correction needed in order to
 estimate $L_{\rm bol}$ from the observed luminosity in any specific
 band. On both these issues, huge progress has been made in the last
 few years (see e.g. \cite{mar04}).

In order to progress from the study of BH mass densities to that of BH
mass functions, we need to break the degeneracy between mass and
accretion rate of any given observed AGN. While in most semi-analytic
works this is done by assuming a constant Eddington ratio for all
sources, here I will propose an alternative, physically motivated, method.
In a recent paper \cite{mhd03} it has been
shown that the hard (2-10 keV) X-ray luminosity of an accreting black holes is
related to its mass and its core radio (at 5GHz) luminosity by the following relation
(the ``fundamental plane'' of
black hole activity):
$\log L_{\rm R}=(0.60^{+0.11}_{-0.11}) \log L_{\rm X}
+(0.78^{+0.11}_{-0.09}) \log M + 7.33^{+4.05}_{-4.07}$;
which can be inverted to  relate BH masses 
to observed nuclear radio and X-ray luminosities:
$ \log M  \simeq  g(\log L_{\rm R}, \log L_{\rm X})$.
One of the consequences of this relation is that, in an ideal case, 
the {\it conditional radio/X-ray} luminosity
function of active black holes, i.e. the number of sources per unit
co-moving volume 
per unit logarithm of radio and X-ray luminosity, 
$\Psi_{\rm C}(L_{\rm R},L_{\rm X})$,
 could be used to reconstruct the mass function of the underlying 
black hole population. 
In fact, the current lack of the exact knowledge of $\Psi_{\rm
  C}(L_{\rm R},L_{\rm X})$ can be (at least partially) 
superseded, given the two separate radio,
$\phi_{\rm R}(L_{\rm R},z)$,  and X-ray, $\phi_{\rm X}(L_{\rm X},z)$,
luminosity functions at redshift $z$, and an {\it independent} estimate of the 
black hole mass function, $\phi_{\rm M}(M,z)$ at the same redshift. 
By taking into account the fundamental plane relationship, we have
that, at any $z$, the conditional luminosity function $\Psi_{\rm C}$ has to 
satisfy the following integral constraints:

\begin{equation}
\label{eq:int_ij}
\phi_{i}(L_{i}) d\log L_{i}=\int_{L_{j,{\rm min}}}^{\infty} \Psi_{\rm
  C}(L_{i}, L_{j}) d\log L_{j}\;\;\;\;\;\;\;(i,j)=({\rm R,X})
\end{equation}

\begin{equation}
\label{eq:int_m}
\phi_{\rm M}(M) d\log M= \int\!\!\!\int_{\log M<g<\log M+d\log M} 
\!\!\!\!\!\!\!\!\!\!\!\!\!\!\!\!\!\!\!\!\!\!\!\!\!\!\!\!\!\!\!\!\!\!\!
\!\!\!\!\!\!\!\!\!\!\!\!\!
\Psi_{\rm C} (L_{\rm X}, L_{\rm R}) d\log L_{\rm R} d\log L_{\rm X}.
\end{equation}

Given observational estimates of $\phi_{\rm X}$, $\phi_{\rm
  R}$ and $\phi_{\rm M}$, 
we start with an initial guess for $\Psi_{\rm C}$, and proceed via
successive iterations, minimizing the
differences between the projections of the conditional luminosity 
function onto the X-ray and radio luminosity axes and the observed
luminosity functions, until a conditional LF is obtained simultaneously
satisfying eqs.~(\ref{eq:int_ij}) and (\ref{eq:int_m}).
Once such an estimate of  $\Psi_{\rm C}$ is found, 
it is possible to derive the local distribution of the X-ray to
  Eddington ratio, and from this, given an appropriate bolometric
  correction (i.e. a specific functional form  $L_{\rm X}=L_{\rm
  X}(M,\dot m)$, where  $\dot m \equiv \epsilon \dot M c^2 / L_{\rm
  Edd}$), the desired accretion rate function.

The redshift evolution of the SMBH population can then be
computed integrating {\it backwards} the continuity equation that
describes SMBH evolution driven by accretion only  \cite{evo}:
\begin{equation}
\label{eq:continuity}
\frac{\partial \phi_{\rm M}(M,t)}{\partial t}+\frac{\partial
[\phi_{\rm M}(M,t) \cdot \langle\dot M(M,t)\rangle]}{\partial M}=0,
\end{equation}
where the mean accretion rate as a function of black hole mass and
time, $\langle \dot M\rangle$ is calculated directly from the
accretion rate distribution function at time $t$. Starting from $z=0$,
the local BHMF, as determined 
independently from the galaxy velocity
dispersion distribution and the $M-\sigma$ relation, 
can be used as a boundary condition to
integrate eq.~(\ref{eq:continuity}) up to the redshift where hard
X-rays and radio luminosity functions of AGN 
can be reliably estimated. The only parameters
 needed are the accretion efficiency
$\epsilon$, and the functional form $L_{\rm X}=L_{\rm X}(M,\dot m)$.

\begin{figure}[t]
\begin{center}
\begin{tabular}{cc}
\includegraphics[angle=270,width=.49\textwidth]{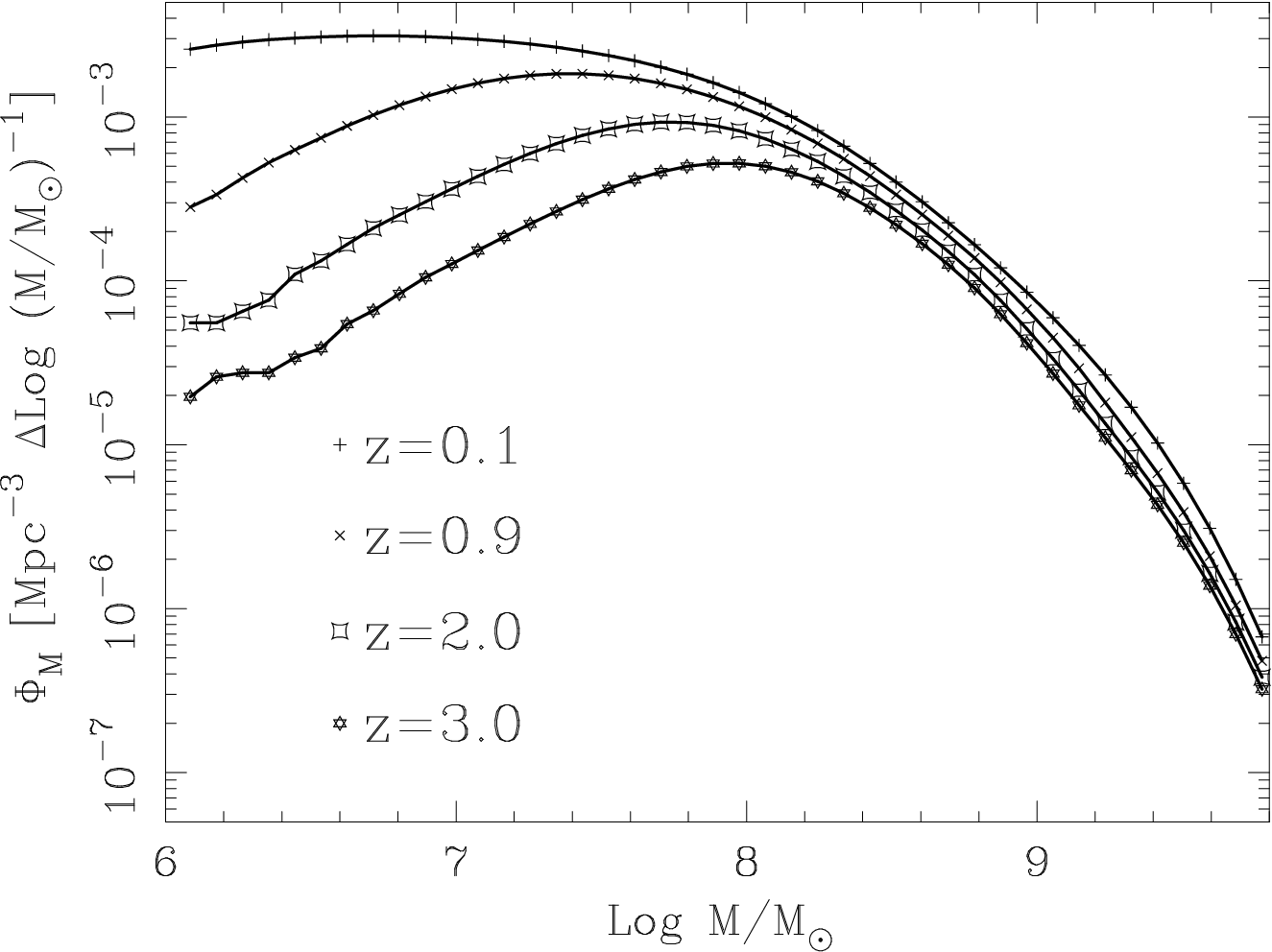}&
\includegraphics[angle=270,width=.49\textwidth]{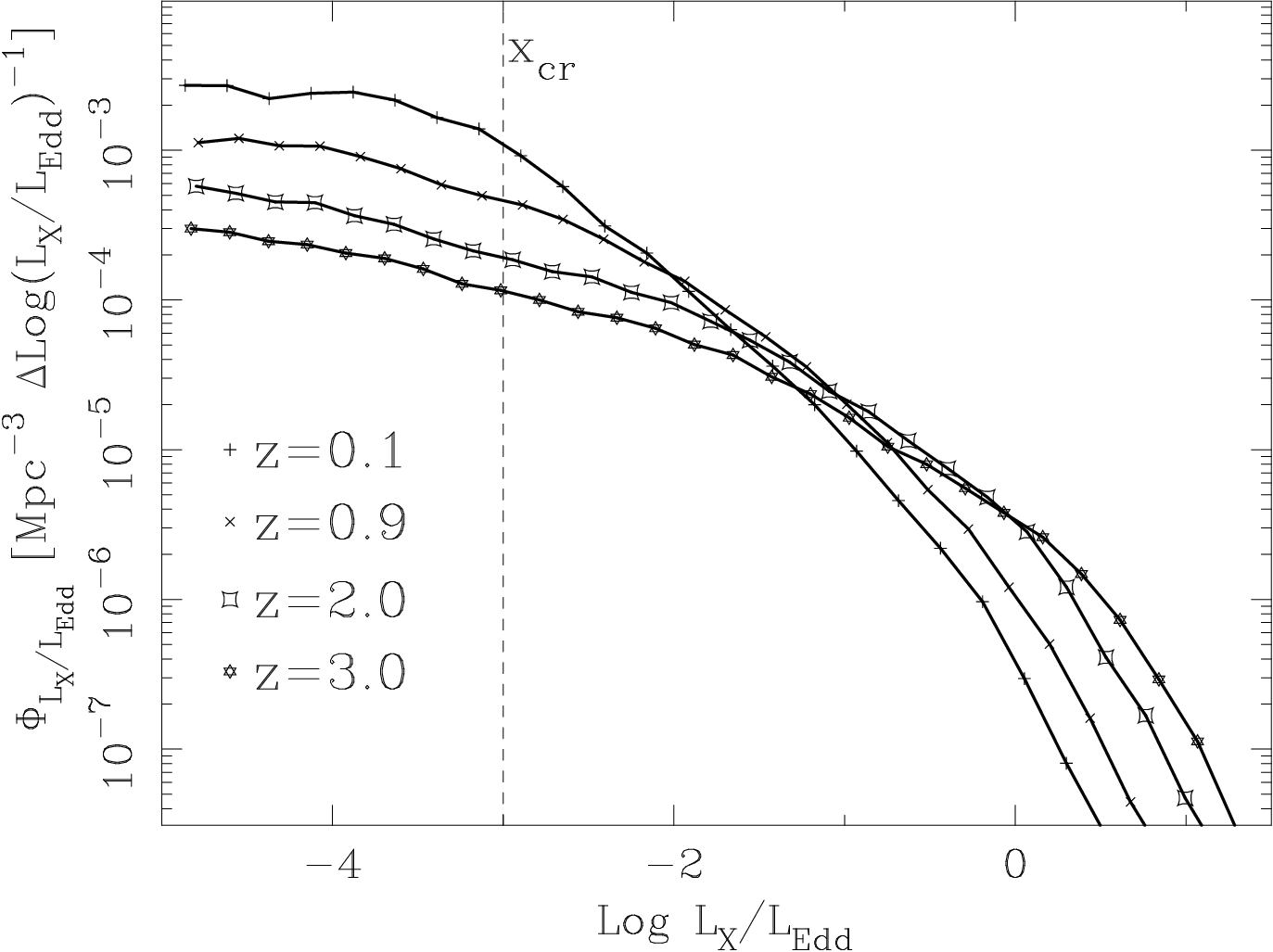}\\
\end{tabular}
\caption{Left Panel: Redshift evolution of the SMBH mass function (BHMF), from redshift
  3 (lower curve) till redshift 0.1 (upper curve). Right Panel:
  Redshift evolution of the SMBH accretion rate function (expressed
  here in terms of X-ray to Eddington luminosity ratio),  
from redshift  3 till redshift 0.1; The vertical dashed line marks the
  adopted value of the critical accretion rate where a transition
  occurs between radiatively inefficient (below) and efficient (above)
  accretion.}
\label{fig:mf}
\end{center}
\end{figure}

Here I present the results obtained using the most recent
determinations of
the hard X-rays and radio LF \cite{ued03,wil01}, fixing $\epsilon=0.1$,
and assuming for the function $L_{\rm X}(\dot m)$ a simple broken
power-law expression,
with $L_{\rm X} \propto \dot m^2$ at low accretion rates (radiatively
inefficient regime) and  $L_{\rm X} \propto \dot m^{0.8}$ at high
luminosities (radiatively efficient regime, see \cite{mar04,mer04} for
details), with the transition being
placed at $x_{\rm cr}=10^{-3}$ \cite{mac03}.  

The evolution of the black hole
mass function is shown in the left panel of Figure~\ref{fig:mf}. 
As opposed to the standard picture of
hierarchical mass build up of dark matter halos in CDM cosmologies,
supermassive black holes growing by accretion between $z\sim 3$ and
now have a mass function
which is more and more dominated by larger mass objects at higher 
redshift. Thus, most of the more massive black
holes ($M > 10^9$) were already in place at $z \sim 3$, and
as a result, the ``typical'' SMBH mass decreases with decreasing redshift.
SMBH appear to be growing  in anti-hierarchical fashion \cite{mar04,gra04}. 

The right panel of Figure~\ref{fig:mf} shows instead the redshift
evolution of the derived accretion rate function (expressed here as
X-ray to Eddington ratio, $L_{\rm  X}/L_{\rm Edd}$). 
While the number of sources accreting at low
rates increases monotonically with decreasing redshift, the situation
is different for rapidly accreting objects, that 
increase in number with increasing redshift. The cut-off redshift,
above which the number of sources declines again, is a function of the
typical X-ray to Eddington ratio, being lower for lower accretion rate 
sources. Once again, it is clear that the ``typical'' X-ray to
Eddington ratio, which can be approximately identified with the knee
of the accretion rate functions of Fig.1, decreases with decreasing
redshift, crossing the critical rate that separates radiatively
inefficient from efficient regimes (here fixed at 10$^{-3}$, vertical dashed line) around
$z \sim 0.5$.

\section{Lifetimes of Active Black Holes}
\label{sec:times}
In all models that derive the properties of the SMBH population
from the observed QSO evolution, a key element is represented by the
typical quasar lifetime or by the almost equivalent activity duty
cycle. 
However, the significance of these parameters is limited to the
standard case in which, on the basis of an observed luminosity
function in a specific waveband, one tries to derive the distribution
of either BH masses or accretion rates. Usually, a constant Eddington
ratio is assumed in this case, which implies that QSO are considered
as on-off switches. Then, the duty cycle is simply the fraction of
black holes active at any time, and the lifetime is the integral of
the duty cycle over the age of the universe (see e.g. \cite{mart03}
and references therein). 

The picture discussed here is different, in that a broad distribution
of Eddington rates is not only allowed, but actually calculated for
the SMBH population at every redshift. When this is the case, a more
meaningful definition of activity lifetime is needed. Let us first define
the mean Eddington rate for object of mass $M_0$ at redshift $z=0$
$\langle \dot m (M_0,z) \rangle$
and then introducing the {\it mean accretion weighted lifetime} of
a SMBH with a given mass {\it today}: 
$\tau(M_0,z) = \int_{\infty}^{z} \langle \dot m (M_0,z') \rangle 
\frac{dt}{dz'} dz'$.
The ratio of $\tau(M_0,z)$ to the Salpeter time, $t_{\rm S}=\epsilon 
M c^2/L_{\rm Edd}= (\epsilon/0.1) 4.5 \times 10^7$ yrs, 
gives the mean number of $e$-folds of mass growth for
objects with mass $M_0$ up to redshift $z$. The ratio of $\tau(M_0,z)$ to the
Hubble time $t_{\rm Hubble}(z)=H(z)^{-1}$, instead, is a measure of the
activity duty cycle of SMBH. 
It is also interesting
here to calculate ``partial'' lifetimes in a given redshift
interval $\Delta z =(z_i,z_f)$:
\begin{equation}
\Delta \tau (M_0,\Delta z) = \int_{z_i}^{z_f} \langle \dot m (M_0,z')
\rangle
\frac{dt}{dz'} dz'.
\end{equation}

\begin{figure}[t]
\begin{center}
\includegraphics[angle=270,width=.70\textwidth]{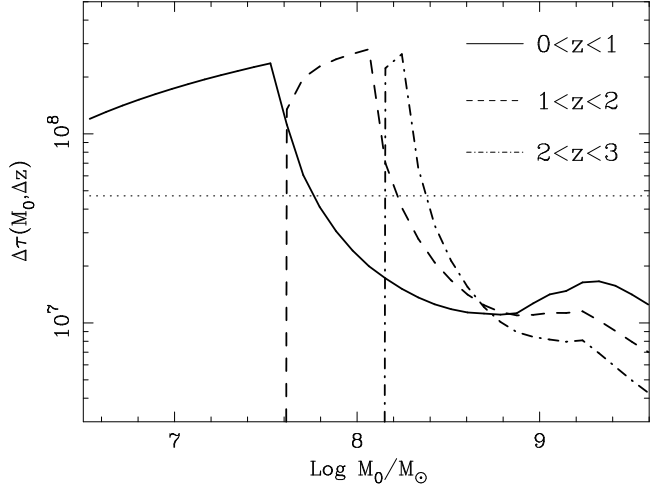}
\caption{Partial mean accretion weighted lifetimes (in years) of SMBH with mass
  today $M_0$, calculated for three different redshift intervals:
  $0<z<1$ (solid line), $1<z<2$ (dashed line) and $2<z<3$ (dot-dashed
  line). The horizontal dotted line is the Salpeter time for accretion
  efficiency of 10\%.}
\label{fig:tau}
\end{center}
\end{figure}

In Figure~\ref{fig:tau}, I show $\Delta \tau (M_0,\Delta z)$ for three
redshift intervals: $0<z\le 1$; $1<z\le 2$ and $2<z\le 3$. The
accretion weighted lifetime for BH of any given mass between $0<z<3$
is of course just the sum of the three. The
anti-hierarchical nature of mass build-up in actively accreting AGN
and QSOs is again clearly illustrated by this plot. In fact, the major
growth episode of a SMBH must coincide with the period when $\Delta
\tau > t_{\rm S}$. This happens at $z<1$ for $M_0 < 10^{7.6}$, between
redshift 1 and 2 for $10^{7.6} < M_0 < 10^{8.2}$, and at $2<z<3$ for $
10^{8.2} < M_0 < 10^{8.4}$. 
Supermassive black holes with masses larger than $M_0 \sim 10^{8.5}$ today,
must have experienced their major episodes of growth at redshift higher than 3.
It also interesting to note that the objects that dominate the SMBH
mass function today, i.e. those in the range of masses around
$10^{7.5}M_{\odot}$, where $M_0
\phi_{M}(M_0,z=0)$ peaks, mainly grew around $z \sim 1$, which
is when most of the X-ray background light we see today was emitted \cite{has03}.

\section{Conclusion}
I have presented a new method to study the growth of accreting
supermassive black holes, based on the simultaneous evolution of the
AGN radio and hard (2-10 keV) X-ray luminosity functions.
The method is based on the locally observed trivariate correlation
between black hole mass, X-ray and radio luminosity (the so-called
fundamental plane of black hole activity, \cite{mhd03}). 
Thanks to this correlation, it is possible for the first time 
to break the degeneracy between luminosity, mass and accretion rate:
QSO and AGN not only grow
in mass during their evolution, but also accrete a different rates
depending on their mass and age.

Qualitatively (i.e. independently on the values of the model
parameters), the evolution of the black hole mass
function between $z=0$ and $z \sim 3$ shows clear signs of an 
{\it anti-hierarchical} behaviour: while the majority of the 
most massive objects
($M > 10^9$) were already in place at $z\sim 3$, lower mass ones
mainly grew at progressively lower redshift, so that 
the average black hole mass increases with
increasing redshift. At the same time, the typical  accretion rate
of SMBH decreases with decreasing redshift. 
Consequently, the lifetime of an actively growing BH, measured through
the partial mean
accretion weighted lifetime (see section~\ref{sec:times}), 
is a strong function of both redshift and black hole mass.

Broadly speaking, what is
presented here is the SMBH analog of the down-sizing of star forming
galaxies. Future more direct and detailed comparisons between SMBH
mass functions, such as those shown here, and galaxy luminosity
functions,  both at low and high redshifts, will hold the key of 
our understanding of the complex interplay
between formation and growth of black holes and galaxies.


\begin{thebibliography}{15.}
\addcontentsline{toc}{section}{References}

\bibitem{cow96} L. L. Cowie, A. Songaila, E. M. Hu, J. G. Cohen: AJ
  \textbf{112}, 839 (1996)

\bibitem{mag} J. Kormendy, D. Richstone: ARA\&A \textbf{33}, 581
  (1995);  J. Magorrian, et al.: AJ \textbf{115}, 2285 (1998);
   L. Ferrarese, D. Merritt: ApJL \textbf{539}, L9 (2000);
   K. Gebhardt, et al.: ApJL \textbf{539}, L13 (2000);
  S. Tremaine, et al.: ApJ \textbf{574}, 740 (2002); 
  A. Marconi, L. K. Hunt: ApJL \textbf{589}, L21 (2003)

\bibitem{hec04} T. M. Heckman, et al.: preprint astro-ph/0406218 (2004)

\bibitem{yt02} A. Soltan: MNRAS \textbf{200}, 115 (1982); 
Q. Yu, S. Tremaine: MNRAS \textbf{335}, 965 (2002)

\bibitem{mar04} A. Marconi, et al.: MNRAS \textbf{351}, 169 (2004)

\bibitem{mhd03} A. Merloni, S. Heinz, T. Di Matteo: MNRAS
  \textbf{345}, 1057 (2003)

\bibitem{evo} T. A. Small, R. D. Blandford: MNRAS \textbf{259}, 725 (1992)

\bibitem{mer04} A. Merloni: MNRAS in press. astro-ph/0402495 (2004)

\bibitem{ued03} Y. Ueda, M. Akiyama, K. Ohta, T. Miyaji: ApJ
  \textbf{598}, 886 (2003)

\bibitem{wil01} C. J. Willott, et al.: MNRAS \textbf{332}, 536 (2001)

\bibitem{mac03} T. Maccarone, E. Gallo, R. P. Fender: MNRAS
  \textbf{345}, L19 (2003)

\bibitem{gra04} G. L. Granato, et al.: ApJ \textbf{600}, 580 (2004)

\bibitem{mart03} P. Martini: in {\it Coevolution of Black Holes and
  Galaxies, from the Carnegie Observatories Centennial
  Symposia} ed. by L. Ho (CUP, Cambridge 2004) p. 170 

\bibitem{has03} G. Hasinger:  AIP Conf.Proc. \textbf{666}, 227 (2003)

\end{thebibliography}
\end{document}